\documentclass[aps,twocolumn,english,superscriptaddress,citeautoscript,showkeys,preprintnumbers,amsmath,amssymb,floatfix,footinbib,prb] {revtex4-2}
\usepackage{graphicx}
\usepackage{soul}
\usepackage{amssymb,amsmath}
\usepackage{hyperref}
\hypersetup{colorlinks=true, linkcolor=red, citecolor=blue}
\usepackage[utf8]{inputenc}
\usepackage[english]{babel}
\usepackage{txfonts}
\usepackage[font=normalsize, caption=false]{subfig}

\usepackage{xcolor}
\usepackage{listings}

\definecolor{vscode-bg}{HTML}{F2F2F2}
\definecolor{vscode-fg}{HTML}{1F1F1F}
\definecolor{vscode-comment}{HTML}{2E7D32}   
\definecolor{vscode-keyword}{HTML}{005CC5}   
\definecolor{vscode-string}{HTML}{A31515}    
\definecolor{vscode-number}{HTML}{986801}    
\definecolor{vscode-function}{HTML}{795E26}  
\definecolor{vscode-class}{HTML}{267F99}     
\definecolor{vscode-operator}{HTML}{1F1F1F}
\definecolor{vscode-line}{HTML}{6E6E6E}
\definecolor{vscode-frame}{HTML}{D0D0D0}

\lstdefinestyle{vscode-python}{
    language=Python,
    backgroundcolor=\color{gray!10},
    basicstyle=\scriptsize\ttfamily,
    keywordstyle=\bfseries\color{vscode-keyword},
    commentstyle=\itshape\color{vscode-comment},
    stringstyle=\color{vscode-string},
    numberstyle=\tiny\color{vscode-line},
    identifierstyle=\color{vscode-fg},
    showstringspaces=false,
    showspaces=false,
    keepspaces=true,
    breaklines=true,
    frame=none,
    numbers=left,
    numbersep=0.5em,
    xleftmargin=1em,
    captionpos=b,
    columns=fullflexible,    
    showtabs=false,                  
    tabsize=2,
    morekeywords={self,None,True,False,as,with,lambda},
    emph={
        set_up_clusters_and_dft, EGQCA, thermodynamic_assessment,
        average_properties, run_ml_relaxation
    },
    emphstyle=\color{vscode-function},
    numberstyle=\tiny\color{black}, 
}

\newcommand{\tc}{$T_\text{c}$}
\newcommand{\tcm}{$T_\text{cm}$}

\newcommand{\nef}{$N_{\varepsilon_\text{F}}$}

\newcommand{\kb}{k_\text{B}}

\begin{document}

\title{EGQCA Code: automated high-throughput thermodynamic modeling of solid solutions}

\author{Dominik Spath}
\email[Corresponding author: ]{dominik.spath@tugraz.at}
\affiliation{Institute of Theoretical and Computational Physics, Graz University of Technology, NAWI Graz, 8010, Graz, Austria}

\author{Luka Wibmer}
\affiliation{Institute of Theoretical and Computational Physics, Graz University of Technology, NAWI Graz, 8010, Graz, Austria}

\author{Kieran Bozier}
\affiliation{Department of Materials Science and Metallurgy, University of Cambridge,
27 Charles Babbage Road, Cambridge CB3 0FS, United Kingdom}

\author{Ralf Meyer}
\affiliation{Institute of Theoretical and Computational Physics, Graz University of Technology, NAWI Graz, 8010, Graz, Austria}

\author{Eva Kogler}
\affiliation{Institute of Theoretical and Computational Physics, Graz University of Technology, NAWI Graz, 8010, Graz, Austria}

\author{Efisio Madeddu}
\affiliation{Dipartimento di Fisica, Sapienza Università di Roma, Roma, Italy}

\author{Chris Pickard}
\affiliation{Department of Materials Science and Metallurgy, University of Cambridge,
27 Charles Babbage Road, Cambridge CB3 0FS, United Kingdom}
\affiliation{Advanced Institute for Materials Research, Tohoku University, Sendai 980-8577, Japan}

\author{Markus Aichhorn}
\affiliation{Institute of Theoretical and Computational Physics, Graz University of Technology, NAWI Graz, 8010, Graz, Austria}

\author{Simone Di Cataldo}
\affiliation{Dipartimento di Fisica, Sapienza Università di Roma, Roma, Italy}

\author{Luiz T. F. Eleno}
\affiliation{Computational Materials Science Group (ComputEEL/MatSci), Universidade de S\~ao Paulo, Escola de Engenharia de Lorena, DEMAR, Lorena, Brazil}

\author{Christoph Heil}
\affiliation{Institute of Theoretical and Computational Physics, Graz University of Technology, NAWI Graz, 8010, Graz, Austria}

\author{Pedro N. Ferreira}
\email[Corresponding author: ]{nunesferreira@tugraz.at}
\affiliation{Institute of Theoretical and Computational Physics, Graz University of Technology, NAWI Graz, 8010, Graz, Austria}

\begin{abstract}
Modeling disordered systems from first principles across the full composition range is a long-standing computational challenge. We present a code package implementing the extended generalized quasichemical approximation (EGQCA), which describes solid solutions through the statistics of a finite set of clusters and enables their thermodynamic modeling at arbitrary compositions and temperatures. 
The package automates the complete workflow and interfaces with both DFT codes and universal machine-learned interatomic potentials (uMLIPs). We show that replacing DFT with uMLIPs reduces the cost of the cluster calculations by up to five orders of magnitude while preserving the thermodynamic predictions. 
To make computationally demanding properties tractable, we additionally introduce a cluster-selection algorithm that identifies the most relevant configurations.
Together, these developments establish EGQCA as a fast and accurate framework for the thermodynamic modeling of solid solutions and a practical route toward high-throughput screening of disordered materials.
\end{abstract}

\date{\today}

\pacs{}

\maketitle

\section{Introduction}
Alloys are indispensable to a wide range of modern technologies, from the automotive~\cite{Zhan2025, Bouaziz2011, DeCooman2018} and aerospace industries~\cite{Pollock2006, Costanza2020} to biomedical implants~\cite{Zhang2019, Nematollahi2019, Rabbitt2024}, the energy sector~\cite{Ran2022, Victoria2001}, and fusion reactors~\cite{Haack2024, Goulatis2024, Wilson2008, Xu2017}. By combining different elements, their properties can be systematically tailored for specific applications, motivating extensive efforts to design new alloys with technological impact~\cite{Akinwekomi_2022, Kolev2026}. Traditionally, alloy design has focused on a few base elements, with small amounts of solutes introduced to modify their properties. More recently, the focus has shifted to alloys containing several principal elements, commonly referred to as high-entropy alloys (HEAs)~\cite{George2019, Schweidler2024, Yeh2004, Cantor2004}, greatly expanding the space of accessible compositions and material properties.

Navigating the vast compositional space of alloys remains tractable as long as the atomic arrangement is ordered.
In such cases, standard density functional theory (DFT) calculations based on small unit cells enable efficient in-silico materials design.
In disordered alloys, however, the constituents randomly occupy the lattice sites, and a faithful description requires large supercells, increasing the computational cost of first-principles predictions significantly. 
Developing efficient methods that treat disorder more realistically therefore remains an active field of research, with multiple existing approaches.

Effective-medium theories, such as the coherent potential approximation (CPA)~\cite{10.1143/PTPS.53.1, PhysRev.175.747, PhysRev.156.1017, PhysRev.156.809} and the virtual crystal approximation (VCA)~\cite{dargam1997critical, bellaiche2000virtual}, replace the random alloy by a single effective medium and thus treat the system within a mean-field picture, neglecting any ordering arising from local correlations. Capturing these microscopic effects requires structural approaches that retain the actual atomic interactions. Among these, cluster-expansion techniques combined with Monte Carlo sampling~\cite{Sanchez1984, Ceder2000, PRXEnergy.3.042001} rely on fitting an effective Hamiltonian to describe the configurational energetics, while the special quasirandom structures (SQS) method~\cite{PhysRevLett.65.353, Wei1990} improves upon random supercell generation by constructing a single cell whose correlation functions reproduce those of the true random alloy at one composition as closely as possible. 

The generalized quasichemical approximation (GQCA)~\cite{sher1987} offers an alternative route to describing alloy thermodynamics across the \textit{full} solubility range. The GQCA is based on representing the crystal as a set of symmetry-inequivalent clusters of fixed size, whose occurrence probabilities are determined by minimizing the free energy of the system, accounting for both mixing enthalpy and configurational entropy. The framework was later extended to incorporate vibrational contributions, leading to the extended GQCA (EGQCA)~\cite{ferreira2024}.
To evaluate the cluster probabilities, only the ground-state energy of each cluster and, optionally, its volume and phonon density of states are required.
Once these probabilities are determined, thermodynamic and physical properties can be readily obtained as cluster-weighted averages at any composition, temperature and pressure, with negligible additional computational cost. 
This provides an efficient framework for capturing the thermodynamic behavior and phase stability~\cite{schleife2010, guedes2019, teles2000, guilhon2017} of disordered alloys and predicting their properties, such as bandgaps~\cite{guilhon2017, mota2024, PhysRevB.73.235205, Bastos2025}, optical absorption~\cite{guilhon2017, Borlido2023, z6kx-pcdr}, superconducting \tc~\cite{ferreira2024}, transport properties~\cite{Borlido2023},  or bulk moduli~\cite{schleife2010, teles2000}, across the full composition range. 
The method has already been successfully applied to a wide range of materials, including semiconductors~\cite{Brivio_GQCA_2016, guedes2019, Borlido2023}, topological insulators~\cite{mota2024}, superconductors~\cite{ferreira2024, Semenok_egqca, lu2026, ferreira2026}, and strongly correlated materials~\cite{wibmer2026}.

With this work, we aim to make the EGQCA method widely available by introducing an open-source, easy-to-use code package that automates the thermodynamic assessment of solid solutions and predicts how their properties evolve with composition.
Beyond these core functionalities, we extend the method to interface with state-of-the-art machine-learned interatomic potentials (MLIPs)~\cite{10.1021/acs.chemmater.9b01294, Batzner2022, Unke_forceFields, batatia2023macehigherorderequivariant, deng2023chgnet, Yu_uMLIPs, 10.1039/d5dd00423c, yang2024mattersim, maceModel, pushing-unconstrained-2026} to reduce the computational cost of ground-state energy  and phonon calculations. We also introduce a cluster-selection algorithm that ranks clusters according to their thermodynamic occurrence probabilities and identifies a small subset of the most relevant configurations. This substantially reduces the number of clusters required to describe computationally costly physical properties, such as electronic spectral functions or electron-phonon coupling.

The method can be applied to investigate phase stability and to identify optimal composition ranges for target material properties. The integration with MLIPs also extends the range of systems accessible to EGQCA. Together, these developments provide a complete workflow for thermodynamic calculations and composition-dependent property evaluation, with direct applications in high-throughput materials screening.
 
The paper is organized as follows. Section~\ref{sec:Theory} introduces the EGQCA formalism for quasi-binary systems. We state the assumptions underlying the approach, derive the governing expressions, and detail the procedure by which the equilibrium cluster probabilities are determined. Section~\ref{sec:code} gives a brief overview of the implementation. In Sec.~\ref{sec:results} the implementation is benchmarked against a chemically diverse set of materials and we show how MLIPs can be used to accelerate the workflow.
We then describe how thermodynamic functions and the composition dependence of physical properties are obtained within the framework, apply the method to calculate the miscibility gap, and conclude by discussing the role of phonons in the EGQCA description.
We summarize our findings in Sec.~\ref{sec:summary} and discuss the computational details in Sec.~\ref{sec:CompDetails}.

\section{EGQCA model}
\label{sec:Theory}
\begin{figure*}[t]
    \includegraphics[width=\linewidth]{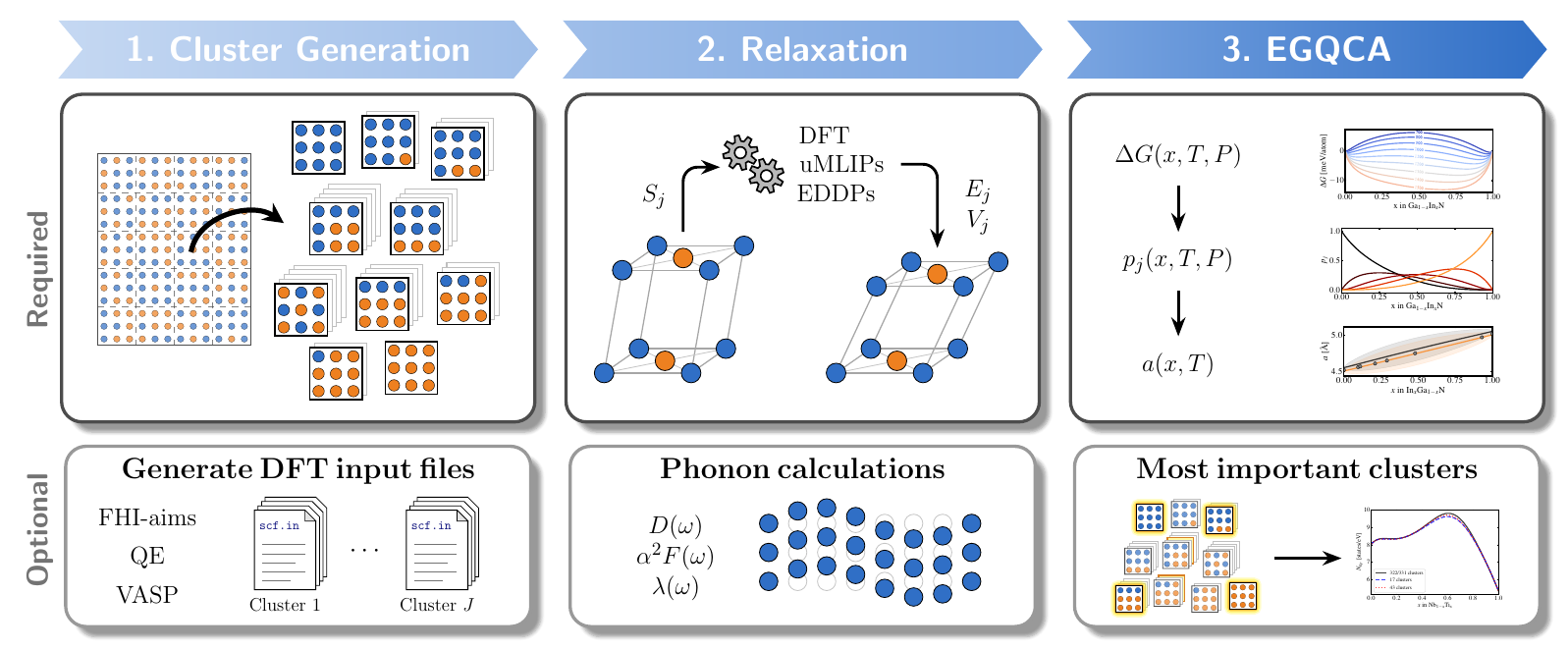}
  \caption{Flowchart of the \texttt{egqca} code. In step (1), all non-equivalent clusters for a given supercell size are generated using the \textsc{supercell} program~\cite{Okhotnikov2016}. The \texttt{egqca} code then organizes each generated CIF file into a dedicated subfolder and automatically produces the corresponding input files for a subsequent DFT calculation, currently supporting FHI-aims, Quantum ESPRESSO, and VASP. In step (2), the ground-state energy of each cluster is computed, either via DFT, uMLIPs or dedicated machine-learning force fields. If the vibrational contribution to the free energy is to be included, the phonon density of states is also calculated at this stage. In step (3), with all required inputs in place, the thermodynamics of the system together with the occurrence probabilities are obtained via the EGQCA method, from which the averaged physical properties can be derived. For computationally demanding quantities, such as the superconducting transition temperature \tc, which may be intractable to evaluate for a large number of clusters, the most relevant clusters can be identified in advance to provide efficient estimates.}
  \label{fig:Flowchart}
\end{figure*}

The GQCA~\cite{sher1987} describes a disordered system by partitioning it into a set of non-equivalent clusters of size $n$.  A solid containing $N$ substitutional sites is thereby represented as an assembly of $M$ clusters, each comprising $n$ atoms.
The current implementation is limited to quasi-binary systems A$_{1-x}$B$_x$C.
For a binary system the number of A and B atoms is fixed by the composition, $N_A=N(1-x)$ and $N_B = Nx$.
The set of non-equivalent clusters, indexed by $j=1,\dots,J$, is generated by permuting the A and B atoms over all sites and assigning a degeneracy $g_j$ to combine symmetry-equivalent clusters. 
Each cluster is characterized by its composition $x_j = n_j/n$, where $n_j$ is the number of B atoms, as well as its energy $E_j$ and volume $V_j$. Since lattice vibrations can significantly affect the thermodynamics of the alloy at higher temperatures, each cluster may additionally be characterized by its vibrational free energy $A_j(T)$~\cite{ferreira2024}. 
A cluster of type $j$ occurs $M_j$ times in the solid solution, allowing its occurrence probability to be defined as
\begin{equation}
    p_j = \frac{M_j}{M}~.
\end{equation}

The equilibrium occurrence probabilities are obtained by minimizing the mixing Gibbs free energy,
\begin{equation}
\Delta G = \Delta H - T\Delta S + \Delta A,
\end{equation}
composed of the mixing enthalpy $\Delta H$, mixing entropy $\Delta S$, and the vibrational free energy of mixing $\Delta A$.
To evaluate $\Delta H$, $\Delta S$ and $\Delta A$, we make two fundamental assumptions: (i) the clusters are energetically decoupled, and (ii) their occurrence is statistically independent except for the fact that they overlap at lattice sites. According to assumption (i), the total energy of a given configuration $\{M_j\}$ reduces to a sum of cluster energies,
\begin{equation}
E(\{M_j\}) = \sum_j M_j E_j ~.
\label{eq:Ham-GQCA}
\end{equation}
For large clusters, such as supercells built by stacking of unit cells, a correction is needed to take into account the overlapping of clusters. In this case, because of assumption (ii) above, each one of the $N$ lattice sites is shared by $n$ clusters, and therefore $E_j$ has to be given per permutable site.

From the statistical independence, an approximate formula for the number of ways to configure the solid solution from a given set $\{M_j\}$ with $N_A/N_B$ atoms follows:
\begin{equation}
W(\{M_j\}) \approx \frac{N!}{N_A!\,N_B!}\,\frac{M!}{\prod_j M_j!}\,\prod_j \left(p_j^0\right)^{M_j}~.
\label{eq:W-GQCA}
\end{equation}
Here, $p_j^0 = g_j\,x^{n_j}(1-x)^{n-n_j}$ is the probability of randomly placing $n_j$ atoms of type B and $n-n_j$ atoms of type A in a cluster of size $n$ and degeneracy $g_j$. 
Applying Stirling's approximation, the configurational entropy follows as
\begin{align}
\Delta S(x,P,T) = \kb \ln W \approx &-N \kb\big[x\ln x + (1-x)\ln(1-x)\big] \nonumber \\
&- M \kb\sum^J_{j=1} p_j \ln\!\left(\frac{p_j}{p^0_j}\right)\,.
\label{eq:S_GQCA}
\end{align}
The first term represents the ideal configurational entropy of a fully random alloy, while the second, expressed as the negative Kullback–Leibler (KL) divergence, accounts for the reduction in entropy arising from deviations of the cluster distribution from the random reference $p_j^0$. Eq.~\eqref{eq:S_GQCA} is a generalization of the Bethe-Peierls entropy~\cite{Bethe_entropy} and, as such, falls within the Cluster Variation Method (CVM) hierarchy~\cite{KikuchiCVM}. A derivation of Eq.~\eqref{eq:S_GQCA} is included in the Appendix.

Historically, the entropy has mostly been approximated using $M=N/n$~\cite{teles2000, ferreira2024, Brivio_GQCA_2016}, which corresponds to non-overlapping clusters.
Setting $M=N$ improves the entropy estimate for overlapping clusters, as it corrects for the double counting of single lattice sites.
This is the case when the clusters are supercells, for which DFT calculations can be performed directly.
When clusters smaller than a unit cell are considered, cluster expansion (CE) interaction energies~\cite{Connolly1983Apr, Sanchez1993Nov} or equivalent methods~\cite{Eleno2003Jan} are necessary, and the correct geometric relationship between $M$ and $N$ must be employed.
Throughout this work, we use supercells as the basic clusters.

The enthalpic contribution is determined from the excess enthalpies of the individual clusters,
\begin{equation}
    \Delta_j = H_j - \frac{n-n_j}{n}H_A - \frac{n_j}{n}H_B ,
\end{equation}
which quantify the deviation of the cluster enthalpy $H_j=E_j+PV_j$ from the composition-weighted linear interpolation of the pure end-members. Here, $E_j$ and $V_j$ are the ground-state energy and volume per permutable site of cluster $j$ at pressure $P$, respectively.
From the excess enthalpies the mixing enthalpy can be obtained:
\begin{equation}
    \Delta H = M\sum^J_{j=1}p_j \Delta_j .
\end{equation}

To evaluate the contribution from the vibrational free energy, we define excess vibrational free energy for each cluster as
\begin{equation}
    \varepsilon_j(T) = A_j(T) - \frac{n-n_j}{n}A_A(T) - \frac{n_j}{n}A_B(T)~,
\end{equation}
where $A_A$ and $A_B$ denote the vibrational free energies per permutable site of the pure A and B
end-members, respectively. Within the harmonic approximation, each $A_j(T)$ is
obtained from the corresponding phonon density of states $D_j(\omega)$ as
\begin{equation}
    A_j(T) = \kb T\int_0^\infty D_j(\omega)\,
    \ln\!\Bigg[2\sinh\!\Big(\frac{\hbar\omega}{2\kb T}\Big)\Bigg]\,d\omega~,
    \label{eq:HarmonicPhDos}
\end{equation}
and the associated contributions to the internal energy and entropy follow from the
standard harmonic expressions~\cite{Korotaev2018}.

Finally, the mixing Gibbs free energy is:
\begin{equation}
\begin{split}
    \Delta G(x,T,P) &= M\sum^J_{j=1} p_j\Delta_j \\
    &+ N \kb T\left(x\ln x + (1-x)\ln(1-x)\right) \\
    &+ M \kb T\sum^J_{j=1} p_j\ln\!\left(\frac{p_j}{p^0_j}\right) \\
    &+ M\sum_{j=1}^J p_j \varepsilon_j(T).
\end{split}
\end{equation}
Minimizing this Gibbs free energy subject to the normalization of the
occurrence probabilities, $\sum_{j}^{J} p_j = 1$, and to the constraint that the
clusters reproduce the target composition, $\sum_{j}^{J} p_j n_j = nx$, results in a polynomial equation:
\begin{equation}
    \sum_{j=1}^J (nx-n_j) g_j e^{-(\varepsilon_j+\Delta_j)\beta}\eta^{n_j} = 0~.
\end{equation}
Here, $\beta = 1/\kb T$ is the inverse temperature and $\eta = x\,e^{\lambda_\mathrm{L}\beta}/(1-x)$, with $\lambda_\mathrm{L}$ being the Lagrange multiplier enforcing the composition constraint. 
Solving this equation yields the Lagrange multiplier, from which the equilibrium probabilities follow:
\begin{equation}
    p_j = \frac{g_j\,\eta^{n_j}\,e^{-(\varepsilon_j+\Delta_j)\beta}}
    {\sum_{i=1}^{J} g_i\,\eta^{n_i}\,e^{-(\varepsilon_i+\Delta_i)\beta}}~.
\end{equation}
The vibrational contribution thus enters the cluster statistics on the same 
footing as the configurational energetics, through the combined exponent $\varepsilon_j + \Delta_j$.
Note that setting $M=N$ in the entropy formula requires the excess free energies to be expressed per permutable site.
From the equilibrium cluster probabilities, the thermodynamic properties of the solid solution can be obtained at the corresponding composition and temperature.

Beyond the mixing thermodynamics, EGQCA allows the evaluation of physical properties for which cluster-resolved quantities are available. Any cluster-resolved property $\mathcal{P}_j$ is averaged over the equilibrium cluster distribution,
\begin{equation}
    \mathcal{P}(x,T,P) = \sum^J_{j=1} p_j(x,T,P)\,\mathcal{P}_j~,
\end{equation}
and the spread of that property across the cluster ensemble is quantified by its
standard deviation:
\begin{equation}
    \Delta\mathcal{P}(x,T,P) = \sqrt{\sum^J_{j=1} p_j(x,T,P)\,\mathcal{P}_j^2
    - \mathcal{P}(x,T,P)^2}~.
\end{equation}
As the weights $p_j$ carry the full composition, temperature, and pressure
dependence, a single set of precomputed cluster quantities $\{\mathcal{P}_j\}$ suffices to map out any such property over the entire phase diagram.

\section{Code workflow}
\label{sec:code}
The equations discussed above are implemented in the \texttt{egqca} code package~\cite{egqca_package}, 
together with a full workflow for generating and setting up the underlying cluster data. 
To obtain all symmetrically inequivalent clusters, the package interfaces with the
\textsc{supercell} program~\cite{Okhotnikov2016}, organizes the CIF file for each cluster
into a dedicated subfolder, and generates the corresponding input files for
FHI-aims~\cite{Blum2009-fhiaims, Lehtola2018}, Quantum ESPRESSO~\cite{QE1, QE2}, and VASP~\cite{kresse1996software, kresse1996efficiency,
PhysRevB.53.7334}. 
Once the cluster energies have been obtained, the EGQCA minimization itself can be executed, as depicted in Fig.~\ref{fig:Flowchart}.

Within the EGQCA approximation, the thermodynamic properties of solid solutions can be evaluated across the \emph{entire} composition range $x \in [0,1]$ at specified temperatures and pressures. The method requires only the cluster energies as input, with cluster lattice parameters and phonon densities of states optionally included to account for additional contributions to the thermodynamics. These quantities can be readily obtained from standard density functional (perturbation) theory calculations or, at a fraction of the computational cost, from universal machine-learned interatomic potentials (uMLIPs) and system-specific potentials such as Ephemeral Data-Derived Potentials (EDDPs)~\cite{EDDP_Pickard, Salzbrenner2023}. From these cluster properties, the thermodynamic behavior across the full solubility range can be reconstructed using the cluster statistics derived above.

In the \texttt{egqca} package, the calculation reduces to initializing the model with the cluster data and evaluating it over a grid of compositions at the desired temperatures:

\begin{lstlisting}[style=vscode-python, language=Python]
from egqca import EGQCA
import numpy as np

m = EGQCA(
    datafile="clusters.dat",
    volume_file="volume.dat",           # optional
    pressure_GPa=10.0,                   # optional
    phonon_file_name="phonons.dos",     # optional
)

m.thermodynamic_assessment(
    np.arange(1e-6, 1.0, 1e-3), 
    Tlist=[600, 800, 1000, 1400]
    )
\end{lstlisting}
The \texttt{EGQCA} constructor method requires only the cluster list (\texttt{clusters.dat}), while optional arguments enable additional contributions to the free energy. Setting a finite \texttt{pressure\_GPa} includes the $PV$ term, evaluated from the cluster lattice parameters provided in \texttt{volume.dat}, whereas \texttt{phonon\_file\_name} incorporates vibrational contributions obtained from the phonon density of states of each cluster. Once initialized, the model retains all cluster data, allowing subsequent analyses to be performed through methods of the same object. The main method, \texttt{thermodynamic\_assessment}, determines the equilibrium cluster probabilities across the specified composition grid for each temperature in \texttt{Tlist}, evaluates the mixing free energy, identifies miscibility gaps, and saves the resulting figures and data files in a dedicated output folder. 
Similarly, \texttt{average\_properties} reads \texttt{properties.dat} and averages each column separately at each temperature:

\begin{lstlisting}[style=vscode-python, language=Python]
m.average_properties(
    "properties.dat",
    x=np.arange(1e-6, 1.0, 1e-3),
    Tlist=[600, 800, 1000, 1400],
    )
\end{lstlisting}
Further details on the input formats and the complete workflow are provided in the Supplemental Material (SM)~\cite{SM} and the package documentation~\cite{egqca_Docs}.

The computational efficiency of EGQCA stems from the fact that the properties of each cluster need to be calculated only once, after which the thermodynamics of the solid solution can be obtained through inexpensive statistical averaging over the cluster ensemble. The computational cost is therefore dominated by the initial evaluation of the cluster ground-state energies and the physical properties $\mathcal{P}_j$ of interest. 
Both limitations in the preceding calculations can be substantially alleviated: the former by replacing DFT calculations with MLIPs, and the latter by reducing the number of clusters for which computationally expensive physical properties must be calculated. In the following section, we explore both strategies.

\section{Benchmark}
\label{sec:results}

\subsection{Machine-Learning-Assisted EGQCA}
When interest is limited to thermodynamic properties, only the ground-state energies of the relaxed clusters are required. Yet even these calculations become infeasible with DFT once the number of clusters reaches into the hundreds. MLIPs provide an alternative to DFT,  reducing the time required for an individual energy evaluation from hours to seconds. In high-throughput settings, universal MLIPs are particularly attractive, as they require no system-specific training,  while reaching accuracies comparable to DFT~\cite{batatia2023macehigherorderequivariant, deng2023chgnet, chen2022universal,merchant2023scaling, yang2024mattersim, riebesell2023matbench, loew2025universal}.

The field of uMLIPs is rapidly evolving, with foundational models improving continuously in both accuracy and transferability. To reflect this diversity, we benchmark three representative models — MatterSim-v1.0.0-5M~\cite{yang2024mattersim}, MACE-MPA-0-medium~\cite{maceModel}, and PET-OMATPES-L~\cite{pushing-unconstrained-2026} — chosen to span different architectural designs, training set compositions, and underlying DFT exchange-correlation functionals. Additional models were tested and produced broadly similar results for the systems considered, although a comprehensive comparison of available uMLIPs is beyond the scope of this work.

To benchmark the uMLIP-accelerated workflow, we draw on known solid solutions and systems to which EGQCA has recently been applied across a broad range of material classes. To ensure that the conclusions are not limited to a particular material class or cluster complexity, the selected set is deliberately diverse, encompassing intermetallics, semiconductors, and insulators, described by both small and larger cluster sizes as well as systems for which phonon calculations are available as an additional benchmark. 

\begin{figure*}[t]
    \includegraphics[width=\linewidth]{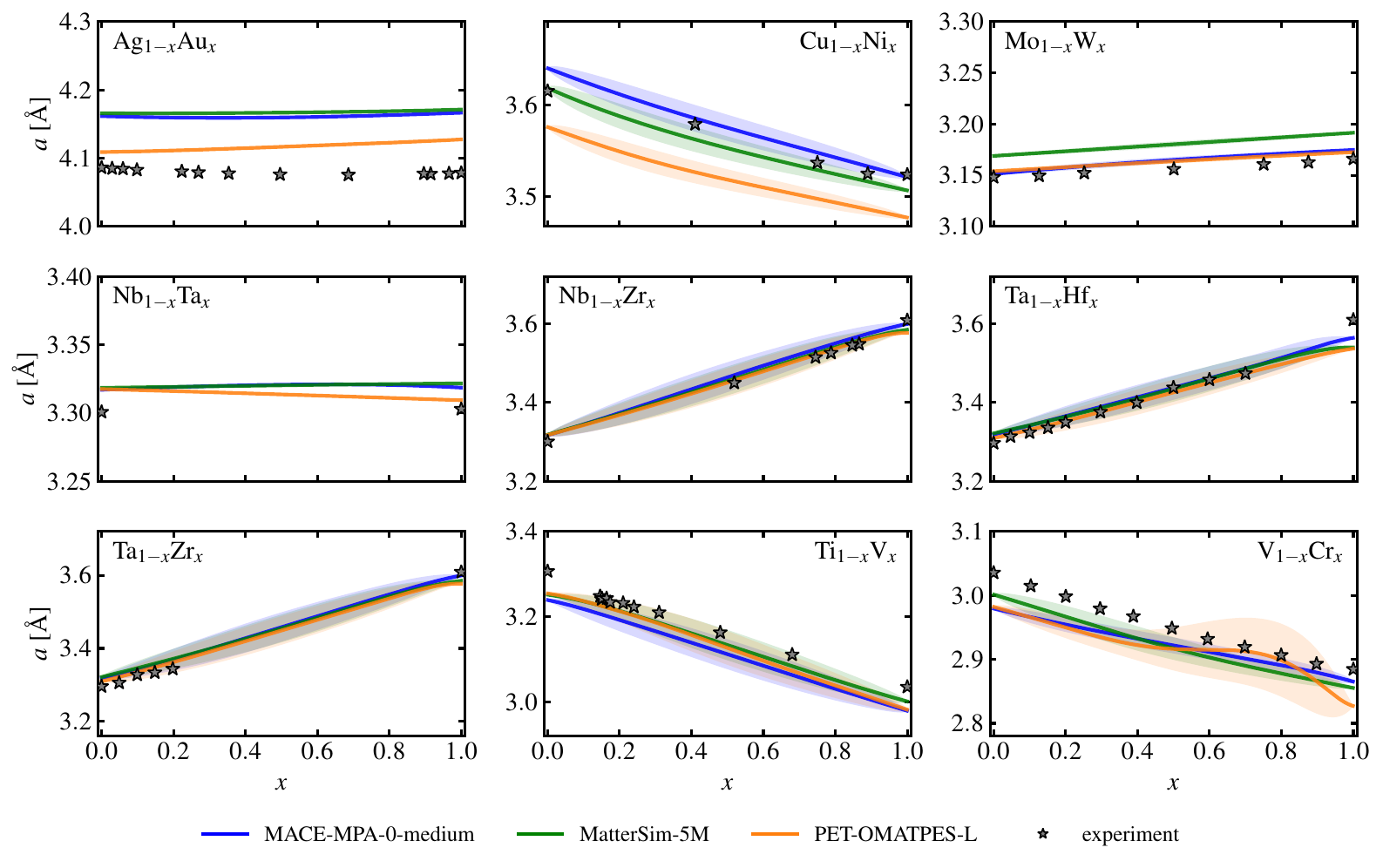}
  \caption{High-throughput screening of the lattice parameters of several solid solutions. Computed with MACE-MPA-0-medium, MatterSim-5M and PET-OMATPES-L with comparison to experiment. Ag$_{1-x}$Au$_x$ and Cu$_{1-x}$Ni$_x$ are in the FCC phase, all other materials are in the BCC phase. For each material, the EGQCA average (solid line) and standard deviation (shaded area) were taken at 1300 K. Experimental data taken from Ag$_{1-x}$Au$_x$~\cite{Bozzolo2007}, Cu$_{1-x}$Ni$_x$~\cite{Kudryavtsev2016}, Mo$_{1-x}$W$_x$~\cite{OhserWiedemann2013}, Nb$_{1-x}$Ta$_x$~\cite{Edwards1951, Grad1995}, Nb$_{1-x}$Zr$_x$~\cite{Grad1995, Skinner1953}, Ta$_{1-x}$Hf$_x$~\cite{PhysRevMaterials.7.064802, Edwards1951, ROSS1963258}, Ta$_{1-x}$Zr$_x$~\cite{PhysRevMaterials.7.064802, Edwards1951, Skinner1953}, Ti$_{1-x}$V$_x$~\cite{Aurelio2002, EPPELSHEIMER1950}, V$_{1-x}$Cr$_x$~\cite{Carlson1959} and were digitized when necessary using PlotDigitizer (https://plotdigitizer.com).}
  \label{fig:HighThroughput}
\end{figure*}

As a first test case, we computed in a high-throughput manner the lattice parameters of several known intermetallic solid solutions using the foundational models MACE-MPA-0-medium, MatterSim-v1.0.0-5M and PET-OMATPES-L, and compared the results to experiment (Fig.~\ref{fig:HighThroughput}). All three models are in close agreement with the experimental data, achieving a total mean relative error (MRE) below 1\%, averaged over all systems. PET-OMATPES-L performs best on average, although it deviates noticeably for Cu$_{1-x}$Ni$_x$ and develops an unphysical kink in V$_{1-x}$Cr$_x$ near $x=0.7$. MACE-MPA-0-medium and MatterSim-5M agree closely for most systems despite their different architectures and training sets. This comparison also shows that running several models side by side is a useful first check on the reliability of the predictions.
A more detailed breakdown of the MRE of each model is given in the SM~\cite{SM}. 

\begin{figure}[t]
    \includegraphics[width=\linewidth]{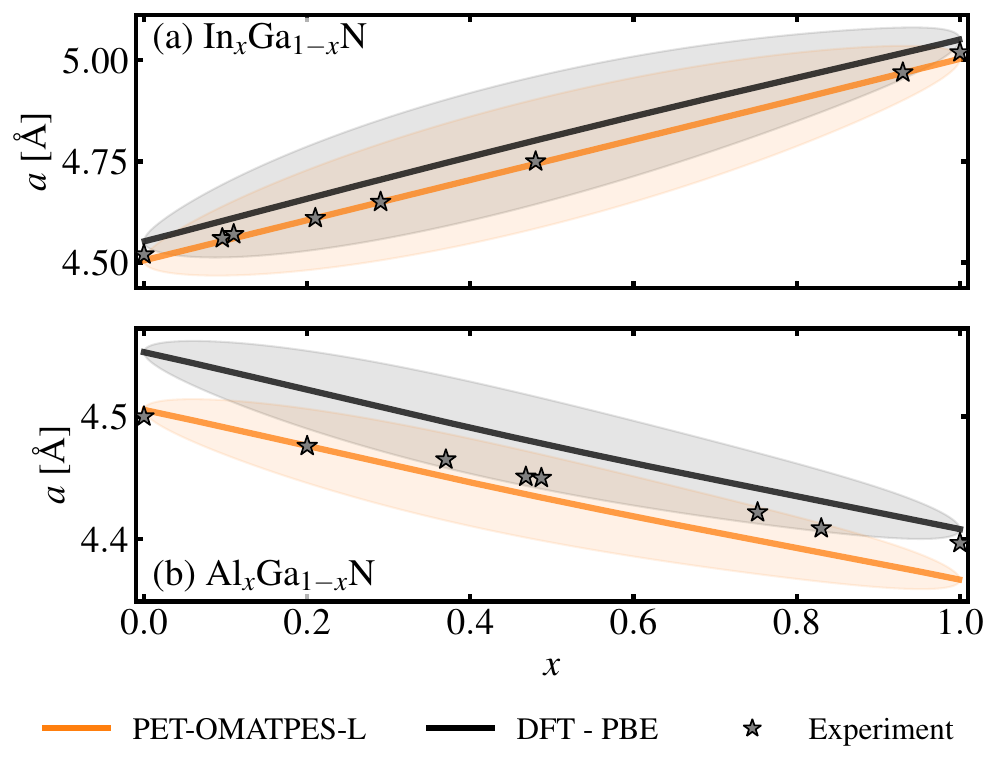}
  \caption{Lattice parameters of Al$_{x}$Ga$_{1-x}$N and In$_{x}$Ga$_{1-x}$N. EGQCA average (solid line) and standard deviation (shaded area) for the lattice parameter of Al$_{x}$Ga$_{1-x}$N and In$_{x}$Ga$_{1-x}$N at $T=950$ K as computed with DFT and the PET-OMATPES-L model. The experimental data for In$_{x}$Ga$_{1-x}$N is taken from Ref.~\cite{garcia_bulk_2015}, that for Al$_{x}$Ga$_{1-x}$N from Ref.~\cite{koizumi_growth_1999} and was extracted using PlotDigitizer (https://plotdigitizer.com).}
  \label{fig:GaN_latpars}
\end{figure}

Having established that the models reproduce the lattice parameters of intermetallic compounds, we now turn to a more demanding case and consider the quasi-binary nitride alloys In$_x$Ga$_{1-x}$N and Al$_x$Ga$_{1-x}$N~\cite{teles2000}. These systems are particularly challenging due to the lattice mismatch between their constituent compounds, especially InN and GaN, which introduces substantial local strain and bond-length distortions. In In$_x$Ga$_{1-x}$N, these effects also lead to pronounced compositional fluctuations and a strong tendency toward phase separation, making the accurate description of their structural properties particularly demanding. Figure~\ref{fig:GaN_latpars} shows the lattice parameters obtained with DFT at the PBE level using FHI-aims~\cite{Blum2009-fhiaims, Lehtola2018} alongside PET-OMATPES-L, with experimental data taken from Refs.~\cite{garcia_bulk_2015, koizumi_growth_1999} for comparison. Both methods reproduce the observed compositional trends, with PET-OMATPES-L showing even closer agreement with the experimental lattice parameters. This difference may be related to the r2SCAN reference data in the training set, as uMLIPs trained on PBE-level data perform comparably to direct DFT-PBE calculations, as shown in the SM~\cite{SM}.

\begin{figure}[t]
    \includegraphics[width=\linewidth]{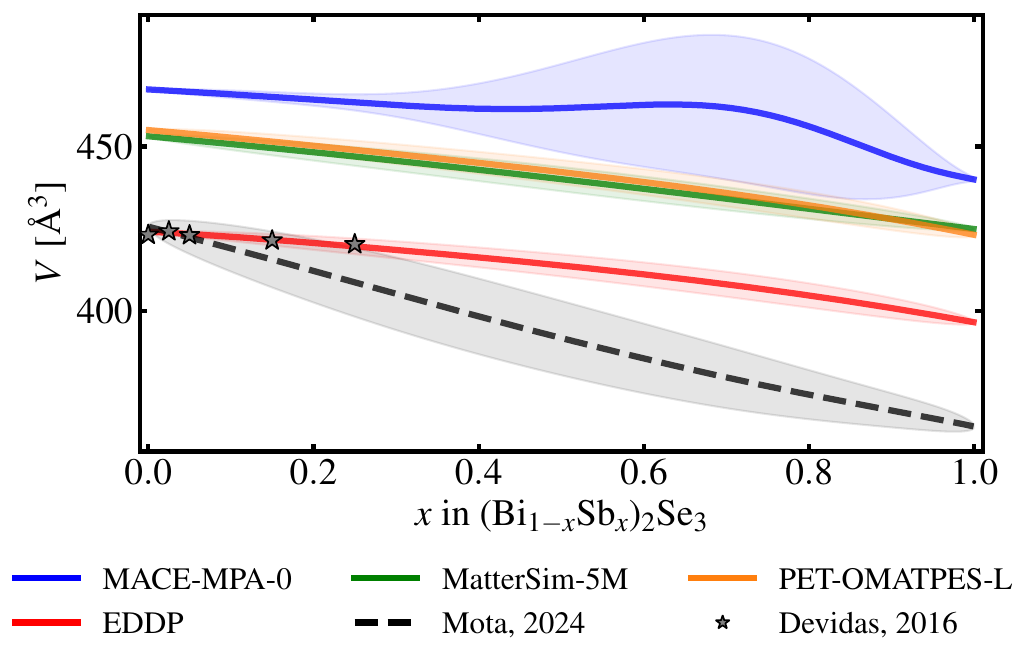}
  \caption{Volume of (Bi$_{1-x}$Sb$_x$)$_{2}$Se$_3$. EGQCA average (solid line) and standard deviation (shaded area) for the volume of (Bi$_{1-x}$Sb$_x$)$_{2}$Se$_3$ at $T=300$ K as predicted by the MatterSim-5M, MACE-MPA-0-medium, PET-OMATPES-L models, and a system-specific EDDP. The predictions are compared to experimental data (Devidas)~\cite{Devidas2017} and to DFT results (Mota)~\cite{mota2024}, for which we recomputed the EGQCA average with our code from the published cluster data.}
  \label{fig:BiSbSe_latpar}
\end{figure}

To further assess the performance of uMLIPs in structurally more complex systems, we consider the layered alloy (Bi$_{1-x}$Sb$_x$)$_2$Se$_3$. Unlike the nitride alloys discussed above, this system presents an additional challenge, as the preferred crystal structure changes with composition, from the rhombohedral structure of Bi$_2$Se$_3$ to the orthorhombic structure of bulk Sb$_2$Se$_3$~\cite{lazcano1999antimony}. Experimentally, the rhombohedral phase is accessible only over a limited composition range. In bulk samples, Sb solubility is restricted to $x\simeq0.32$~\cite{kuznetsov1963equilibrium, Devidas2017}, whereas thin films can be grown up to $x\simeq0.5$~\cite{Matetskiy2020}, with the precise solubility limit depending on the substrate~\cite{Liu2017, Satake2018}.

We employ a $2\times2\times2$ rhombohedral supercell, which preserves the symmetry of the underlying crystal structure while remaining computationally affordable through the use of MLIPs. In Fig.~\ref{fig:BiSbSe_latpar}, we compare the volumes predicted by uMLIPs, and DFT from Mota et al.~\cite{mota2024} against experimental data~\cite{Devidas2017}. For this layered system, the uMLIP predictions deviate substantially from both experiment and the DFT reference, with considerable differences also observed among the individual models. While MatterSim-5M and PET-OMATPES-L reproduce the qualitative compositional dependence, MACE-MPA-0-medium exhibits a pronounced kink in the Sb-rich region, suggesting that its predictions become unreliable at these compositions.

These discrepancies with experiment can be related to the underlying physics of the system. In particular, the largest deviations occur in the compositional region where the rhombohedral phase competes with the orthorhombic structure. Furthermore, the volume of layered materials is strongly influenced by weak van der Waals interactions between the layers, which are not adequately described by models trained on DFT data without non-local corrections. 
The difference between the results of Mota et al. and the uMLIPs can be attributed to their use of the DFT-1/2 method~\cite{Tao2017, Matusalem2021} together with the smaller cluster set derived from the $2\times2\times1$ supercell.

The limitations observed for (Bi$_{1-x}$Sb$_x$)$_2$Se$_3$ reflect a broader challenge associated with the transferability of uMLIPs. Most existing models are trained predominantly on bulk materials near equilibrium, and their predictive accuracy can deteriorate when applied to configurations that are poorly represented in their training sets~\cite{Deng2025, mlipBias_Wong}. Similar caution is therefore warranted for systems outside this domain, including two-dimensional materials, surfaces, structures under pressure, and configurations far from the thermodynamic convex hull. 
Since such limitations are not always apparent a priori, comparing several independent uMLIPs provides a simple first check. Deviating predictions among models are a clear warning sign that the results cannot be trusted without further verification, for example by performing DFT calculations for a representative subset of clusters.

Possible routes to improve the description are to fine-tune a uMLIP or to train an interatomic potential specifically for the system. However, for disordered systems, this remains challenging due to the limited availability of suitable training data and the large number of possible atomic configurations.
Nonetheless, here we trained a dedicated EDDP~\cite{EDDP_Pickard, Salzbrenner2023} on PBE+D4~\cite{Caldeweyher2019, Caldeweyher2020} data to describe the R$\bar{3}m$ structure of (Bi$_{1-x}$Sb$_x$)$_2$Se$_3$ over the entire composition range, with the results shown in Fig.~\ref{fig:BiSbSe_latpar}.
The EDDP reproduces the experimental volumes across the measured range and significantly improves upon the DFT reference of Mota et al.~\cite{mota2024}. 
This improved agreement can be attributed both to the larger cluster set derived from the $2\times2\times2$ supercell, which the EDDP makes computationally accessible, and to training data that explicitly account for van der Waals interactions, which are essential for accurately describing the interlayer separation.

\subsection{Cluster reduction}
One of the main advantages of the EGQCA method is that it allows any physical property to be investigated across the entire composition range. In practice, however, its applicability is limited by the cost of evaluating that property for each cluster.
To overcome this, we exploit the fact that clusters with the same number of substituted atoms often exhibit similar physical properties, allowing the full cluster ensemble to be represented by a much smaller subset of configurations. We developed a dedicated cluster-selection algorithm that ranks clusters according to their occurrence probabilities integrated over the composition range and selects the most relevant configurations at each cluster composition $x_j$. To preserve coverage of the full composition range, at least one cluster per $x_j$ is retained. 

\begin{figure}[t]
    \includegraphics[width=\linewidth]{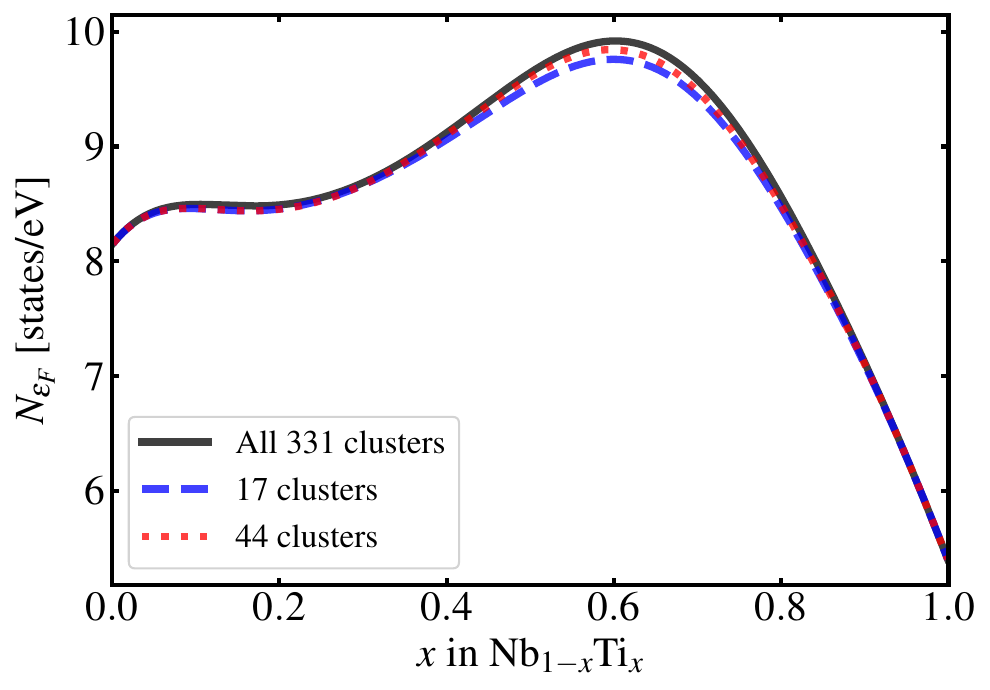}
  \caption{DOS at the Fermi level of Nb$_{1-x}$Ti$_x$. EGQCA average of the DOS at the Fermi level for Nb$_{1-x}$Ti$_x$ under 300 GPa pressure at 800 K for all clusters of the $2\times2\times2$ supercell compared to the reduced subsets of 17 and 44 clusters, respectively. } 
  \label{fig:NbTi_DOS}
\end{figure}

We applied this approach to Nb-Ti in the BCC $\beta$-phase at 300 GPa, where high pressure is necessary to dynamically stabilize the pure Ti reference cluster. 
Ground-state energies and densities of states were obtained from DFT calculations for each cluster, with the DOS at the Fermi level \nef\ serving as a key observable to test the validity of the cluster reduction algorithm. Importantly, clusters sharing the same number of substituted atoms exhibit similar \nef\ values, confirming that the electronic structure of the Nb-Ti system is governed primarily by composition rather than the exact arrangement of atoms within a cluster. 
As shown in Fig.~\ref{fig:NbTi_DOS}, a reduced set of just 17 clusters, out of the original 331, is sufficient to reproduce the compositional dependence of \nef. Increasing the set to 44 clusters yields only marginal improvement.
These results demonstrate the effectiveness of the cluster reduction algorithm for Nb$_{1-x}$Ti$_x$.
Notably, this reduced set can be identified already with uMLIPs. We use MatterSim-5M, whose training data cover pressures up to 1000~GPa~\cite{yang2024mattersim}.
Applying the algorithm to its energies selects largely the same clusters, with 15 of the 17 shared for the smaller subset and 37 of 44 for the larger one, showing that uMLIPs reliably pick out the configurations that dominate the thermodynamics of the system.

\subsection{Miscibility gap}
A key strength of EGQCA is that it provides direct access to the phase stability of solid solutions, allowing us to determine whether an alloy remains homogeneous or separates into phases of different composition.
The central quantity in this context is the miscibility gap, which defines the composition range where a homogeneous solid solution is thermodynamically unfavorable with respect to phase separation. Determining the miscibility gap is essential for materials design, as it identifies the compositions that can be experimentally realized as a single homogeneous phase. However, phase separation can even be beneficial, as it can be exploited to engineer microstructures with improved mechanical properties, including hardness, strength, and ductility~\cite{divilov2024spinodal}. Predicting miscibility gaps is therefore essential for both identifying suitable synthesis conditions and for designing materials in which controlled phase separation is desirable.

The formation of a miscibility gap is governed by the competition between the enthalpy and entropy of mixing. When mixing is energetically unfavorable, a homogeneous solid solution may lower its free energy by separating into two phases of different compositions. At sufficiently high temperatures, the entropy of mixing favors the homogeneous phase, whereas at lower temperatures the enthalpic contribution can dominate, giving rise to a miscibility gap. The stability of the solid solution is therefore determined by the shape of the mixing Gibbs free energy $\Delta G(x,T)$, from which two important boundaries can be identified: the binodal and spinodal lines.

The binodal line defines the equilibrium compositions of the two coexisting phases and is obtained through the common-tangent construction applied to $\Delta G(x,T)$. The spinodal line, in contrast, marks the limit of stability against infinitesimal compositional fluctuations and is determined by the inflection points of the free-energy curve, where $\partial^2 \Delta G/\partial x^2 = 0$. Together, these boundaries divide the composition range into three distinct regimes. Outside the binodal, the homogeneous solid solution is thermodynamically stable. Between the binodal and spinodal, the system is metastable: small compositional fluctuations are unfavorable, but sufficiently large fluctuations can drive phase separation through nucleation and growth. For regions inside of the spinodal line, the free-energy curvature becomes negative, and even infinitesimal fluctuations lower the free energy (ignoring gradient terms), leading to spontaneous phase separation without a nucleation barrier, a process known as spinodal decomposition. As the temperature increases, the binodal line and spinodal line approach each other and eventually meet at the critical temperature $T_\text{cm}$, above which the miscibility gap disappears and the solid solution becomes stable across the entire composition range.

\begin{figure}[t]
    \includegraphics[width=\linewidth]{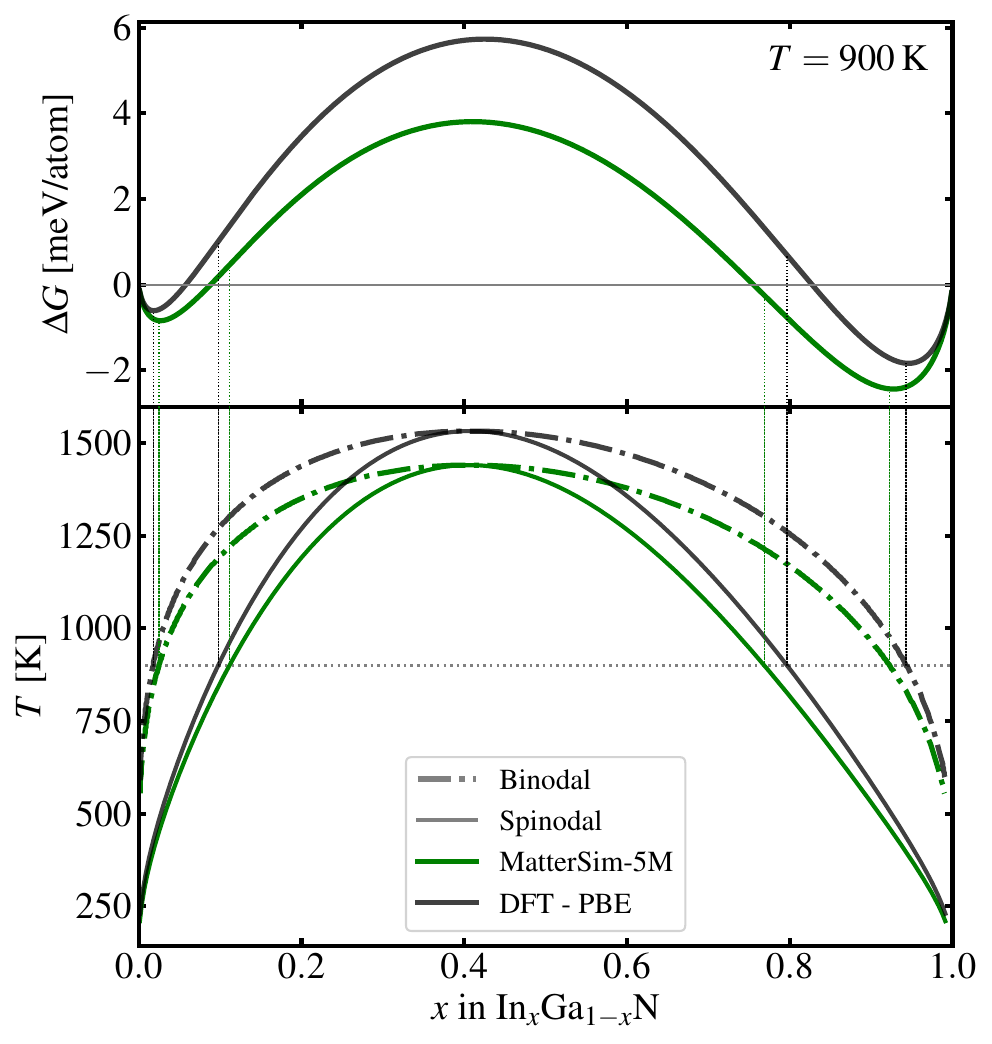}
  \caption{Miscibility gap and free energy at $T=900$ K of In$_{x}$Ga$_{1-x}$N. Calculated from DFT and MatterSim-5M data. The predicted critical temperatures are $T_{\text{cm}}^{\text{DFT}}=1531.9$ K and $T_{\text{cm}}^{\text{MS}}=1440.0$ K for DFT and MatterSim-5M, respectively. The vertical lines indicate the location of the spinodal and binodal points at $T=900$ K.}
  \label{fig:InGaN_miscgap}
\end{figure}

Both boundaries can be determined directly from $\Delta G(x,T)$, which is already obtained as part of the standard thermodynamic assessment in EGQCA. Within the approximation that neglects vibrational contributions, only the ground-state energies of the clusters are required, making the evaluation of miscibility gaps essentially free once these energies are available. Combining EGQCA with uMLIPs therefore provides an efficient route to predicting phase stability, substantially reducing the computational cost associated with generating the cluster data.

To demonstrate this capability, we return to the In$_x$Ga$_{1-x}$N system discussed above. As shown in Fig.~\ref{fig:InGaN_miscgap}, MatterSim-5M reproduces the miscibility gap predicted by DFT at the PBE level remarkably well. It gives the same critical composition, $x_c=0.39$, and the critical temperatures differ by only about 6\,\%. 
The agreement extends even further, as both approaches predict nearly identical shapes for the mixing Gibbs free energy, with MatterSim-5M yielding slightly lower values at intermediate compositions. 
The other two uMLIPs, PET-OMATPES-L and MACE-MPA-0-medium, yield qualitatively similar miscibility gaps as shown in Fig.~5 in the SM.
Importantly, this agreement is obtained at substantially reduced computational cost. MatterSim-5M completes the calculation in just 74.0\,s on a single CPU core of a standard workstation, compared to 1863.6\,core-hours for the full DFT workflow, including convergence tests and structural relaxation. This represents a reduction of nearly five orders of magnitude in computational effort, demonstrating the potential of uMLIPs for high-throughput investigations of alloy thermodynamics.

\begin{figure*}
    \includegraphics[width=1\linewidth]{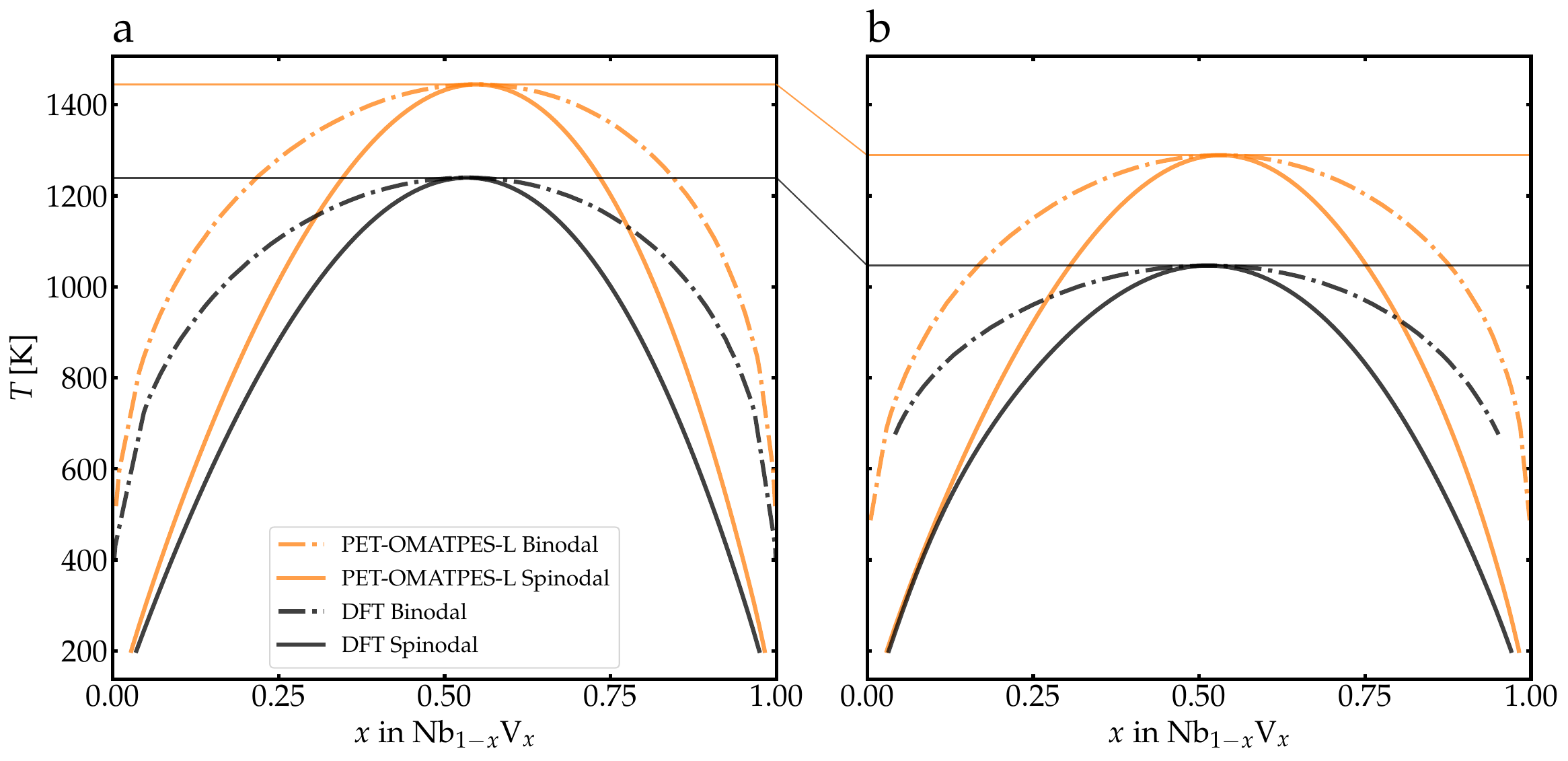}
  \caption{Miscibility gap of Nb$_{1-x}$V$_x$. Calculated with PET-OMATPES-L and compared to DFT~\cite{ferreira2024}, recomputed with the present entropy expression ($M=N$). The left panel (a) shows the miscibility gaps without phonon contribution, while in the right panel (b) phonon contributions are included.}
  \label{fig:NbV_miscibility_gap}
\end{figure*}

While the results for In$_x$Ga$_{1-x}$N show that ground-state energies alone can already yield an accurate description of alloy thermodynamics, lattice vibrations can play an equally important role in determining phase stability at finite temperatures~\cite{RevModPhys.74.11, FULTZ2010247, PhysRevB.53.8993}. In particular, differences in the vibrational free energies of the alloy configurations and their end-members can significantly modify the mixing Gibbs free energy, shifting the miscibility gap and its critical temperature~\cite{PhysRevB.58.R5897, PhysRevB.86.020201}. Accounting for these contributions is therefore essential for a reliable description of many alloy systems.

To address this, the \texttt{egqca} package supports the inclusion of vibrational contributions through Eq.~\ref{eq:HarmonicPhDos}. The phonon DOS can be calculated either with DFT or with MLIPs. For DFT, the \texttt{egqca} package can provide the necessary input files for each cluster, whereas the MLIP approach is entirely included within the package. The implementation builds on the \texttt{ase.Phonons}~\cite{ASE-Lib} class, which calculates the harmonic force constants through finite differences, using the chosen interatomic potential to evaluate the atomic forces. The resulting phonon density of states is then used to calculate the vibrational free energy of each cluster, which enters the mixing Gibbs free energy within the EGQCA framework. This allows the effects of lattice vibrations on phase stability to be incorporated without the computational expense of conventional first-principles phonon calculations.

The importance of vibrational contributions has already been demonstrated within the EGQCA framework for Nb$_{1-x}$V$_x$ using DFT~\cite{ferreira2024}. In this system, calculations based on ground-state energies alone overestimate the experimentally reported critical temperature of the miscibility gap, 1077\,K~\cite{GAO2018316}, by about 15\,\%. Including vibrational contributions substantially lowers the critical temperature, bringing the DFT predictions into agreement with experiment.
We use this system to assess whether uMLIPs can also capture the influence of lattice vibrations on alloy thermodynamics. 

Figures~\ref{fig:NbV_miscibility_gap}a and~\ref{fig:NbV_miscibility_gap}b compare the miscibility gaps calculated with PET-OMATPES-L against the DFT results, without and with vibrational contributions, respectively. In both cases, the uMLIP reproduces the overall shape of the DFT miscibility gap and, importantly, captures the reduction in critical temperature upon including phonons. The resulting prediction of \tcm\ = 1290\,K differs from experiment by approximately $20$\,\%. MatterSim-5M and MACE-MPA-0-medium likewise capture the phonon-induced reduction of \tcm\ (Supplementary Fig.~6).
Together with the In$_x$Ga$_{1-x}$N results, this suggests that selected uMLIPs can reproduce qualitative thermodynamic trends in solid solutions, including changes in phase stability associated with configurational and vibrational contributions.

The combination of EGQCA and uMLIPs opens up possibilities beyond modeling individual solid solutions. The miscibility gaps considered here describe phase separation within a given phase, but the same framework can, in principle, be applied to different competing structures. By evaluating and comparing their Gibbs free energies, it becomes possible to construct complete phase diagrams across composition, temperature, and pressure, including transitions between different crystal structures. The computational efficiency of uMLIPs makes such investigations particularly attractive for high-throughput applications.

\section{Summary}
\label{sec:summary}
We have introduced the \texttt{egqca} code package, an easy-to-use framework for describing the thermodynamics of solid solutions across the full range of solubility. It treats the contributions in disordered systems on an equal footing through the mixing enthalpy, the configurational entropy, and the vibrational free energy.
The entire workflow is integrated into a single package, covering everything from cluster generation and ground-state energy evaluation - either directly through universal machine-learned interatomic potentials or by generating the corresponding DFT input files - to the statistical modeling of physical properties as a function of composition. 

Coupling \texttt{egqca} to universal MLIPs addresses the bottleneck of evaluating ground-state energies when the cluster space grows into the thousands or when the individual clusters themselves become large. 
Beyond the energy evaluation itself, uMLIPs can be useful for pre-relaxing structures or performing convergence tests with respect to cluster size. 
The In$_x$Ga$_{1-x}$N miscibility gap illustrates what this means in practice, with a workflow of well over a thousand core-hours collapsing to under two minutes while the physics remains essentially unchanged.

Our benchmarks show that uMLIPs accurately reproduce composition-dependent properties such as lattice parameters, miscibility gaps, and their vibrational corrections for a broad range of systems. 
For certain  chemically and structurally complex cases, however, the model predictions can deviate from the experimental data. In such cases, the reliability of the results can be assessed by comparing several independent potentials with one another and with DFT calculations for a few representative clusters.
Where uMLIPs fail and DFT is too expensive, MLIPs trained specifically for the system offer a route to retain high accuracy at a manageable computational cost.

Furthermore, the \texttt{egqca} package provides an algorithm that ranks clusters by their integrated probability weight and selects the most relevant ones at each composition. Retaining only these dominant clusters preserves the essential physics while reducing the expansion enough to make demanding observables accessible across the full composition range.
For Nb$_{1-x}$Ti$_x$ under pressure this reduces the set from $331$ to $17$ clusters while still capturing the compositional dependence of \nef.

Taken together, these features make \texttt{egqca} a fast and reliable tool for the thermodynamics of solid solutions over the full solubility range. It applies across chemically diverse systems and gives access to quantities ranging from electronic properties to the thermodynamics of phase stability. Future developments will extend the formalism to multi-component systems and to the explicit modeling of multiple phases, opening a route towards efficiently computed phase diagrams.

\section{Methods}
\label{sec:CompDetails}

\textbf{Nb$_{1-x}$Ti$_x$ DFT calculations:}
The electronic properties of Nb$_{1-x}$Ti$_x$ at 300\,GPa were investigated from first principles in a 2$\times$2$\times$2 supercell. Electronic-structure calculations were performed within DFT~\cite{hohenberg1964, kohn1965} using the Quantum ESPRESSO package~\cite{QE1, QE2}. The electron--ion interaction was described using scalar-relativistic optimized norm-conserving Vanderbilt (ONCV) pseudopotentials~\cite{ONCV1, ONCV2} from the PseudoDojo library~\cite{PseudoDojo}, with exchange and correlation effects treated using the PBE functional~\cite{perdew1996}. The plane-wave kinetic-energy cutoff was set to 75\,Ry, with a corresponding charge-density cutoff of 300\,Ry. Brillouin-zone integrations were performed using unshifted 8$\times$8$\times$8 Monkhorst--Pack $k$-point meshes~\cite{monkhorst1976}, with metallic occupations described by Methfessel--Paxton smearing~\cite{MP-smearing} of 0.01\,Ry. A self-consistency threshold of 10$^{-10}$\,Ry was adopted for the electronic calculations.

Structural optimizations were carried out at a fixed external pressure of 300\,GPa using the BFGS algorithm~\cite{BFGS1, BFGS2, BFGS3, BFGS4}. Both the lattice parameter and internal atomic coordinates were relaxed, while preserving the cubic symmetry of the underlying BCC supercell. To ensure consistent structural parameters, the relaxation workflow employed successive variable-cell optimizations, using the converged lattice parameter and atomic positions from each step as input for the next. These were followed by a fixed-cell relaxation of the internal coordinates and a final self-consistent electronic calculation using the optimized structure.

\textbf{In$_x$Ga$_{1-x}$N and Al$_x$Ga$_{1-x}$N DFT calculations:}
Density functional theory calculations were performed with the all-electron, atom-centered orbital code FHI-aims ~\cite{Blum2009-fhiaims, Lehtola2018} (version 250822.1), using the PBE-GGA exchange and correlation functional~\cite{perdew1996} and the atomic ZORA scalar-relativistic treatment. The 2020 intermediate species defaults were used for the basis sets and integration grids of Al, In, Ga and N. Calculations were spin-unpolarized. The clusters were derived from the 8-atom conventional cell of zinc-blende GaN (space group F$\bar{4}$3m), giving five symmetry-inequivalent configurations. The Brillouin zone was sampled with a k-point density of 4~\AA{} along each reciprocal lattice vector. A Gaussian occupation smearing of 0.1 eV was applied. Both the k-point density and the smearing were checked for convergence on the GaN and InN end members. Self-consistency thresholds were 10$^{-9}$ eV for the total energy, 10$^{-3}$ eV for the sum of eigenvalues and 10$^{-8}$ for the charge density. Core states below $-200$ eV were kept frozen during the SCF cycle. Lattice parameters (with fixed cell angles) and atomic positions were relaxed with the trust-radius method using analytical forces and stress, until all residual forces were below $5\times10^{-3}$ eV/Å.

\textbf{EDDPs:} Ephemeral data-derived potentials (EDDPs) combine a generalized Lennard-Jones descriptor with a lightweight neural network~\cite{EDDP_Pickard, Salzbrenner2023}. Here, the potential was trained on DFT energies computed with \textsc{Castep}~\cite{clarkFirstPrinciplesMethods2005} using PBE+D4, a plane-wave cutoff of 340~eV, QC5 pseudopotentials, and a $\mathbf{k}$-point spacing of $2\pi \times 0.05$~\AA$^{-1}$.  The initial training data consisted of 3160 randomly perturbed supercells of up to 20 atoms, based on the $\mathrm{R}\bar{3}m$ Bi$_2$Se$_3$ structure and with substitution of Sb for Bi along the (Bi$_{1-x}$Sb$_x$)$_2$Se$_3$ tie line. The potential was then refined with five cycles of active learning, in which EDDP minima were evaluated with DFT and added to the training set, resulting in a total of 3500 structures. This final potential was then used to model the 40-atom supercells used within the EGQCA.

\textbf{uMLIPs:} All structural relaxations with universal machine-learned interatomic potentials (uMLIPs) were performed with the \texttt{egqca} Python package~\cite{egqca_package}, which interfaces the uMLIPs through the Atomic Simulation Environment (ASE, version 3.28.0)~\cite{ASE-Lib} and automates the relaxation of all clusters. We used  MACE-MPA-0-medium~\cite{maceModel} via mace-torch 0.3.15, MatterSim-v1.0.0-5M ~\cite{yang2024mattersim} via mattersim 1.2.3, and PET-OMATPES-L~\cite{pushing-unconstrained-2026} via upet 0.2.2.  Atomic positions and the cell were relaxed together using the FrechetCellFilter at zero external pressure, except for Nb-Ti where 300 GPa was used. The FIRE optimizer was used with a convergence criterion of $f_\mathrm{max}$ = 10$^{-4}$ eV/\AA{} on the atomic and cell degrees of freedom. During relaxation, the cell metric was constrained to the Bravais lattice of the parent structure, while the atoms relaxed freely. 

\textbf{EGQCA:} All EGQCA calculations were performed with the \texttt{egqca} Python package~\cite{egqca_package}. It generates the clusters, interfaces the DFT codes (FHI-aims, Quantum ESPRESSO and VASP) and the uMLIPs for the cluster relaxations, collects the cluster energies and lattice parameters, and carries out the thermodynamic assessment.  Symmetry-inequivalent configurations of the substitutional sublattice, together with their degeneracies $g_j$, were generated with the Supercell program~\cite{Okhotnikov2016}. 

\section*{Appendix: the GQCA Entropy expression}
\label{eq:Appendix}










As usual, we will use Boltzmann equation for the entropy $S$ of a microcanonical ensemble,
\begin{equation}
    S = k_B \ln W
\end{equation}
in which $k_B$ is Boltzmann constant and $W$ is the number of  microstates, or microscopic configurations, in which the components (i.e., atomic species, for instance) of the system can be arranged.

As described in the main paper, $N$ is the number of (permutable) sites in the lattice, each occupied by one of two possible species, $A$ and $B$. 
The total number of each species will be indicated by $N_A$, and $N_B$, respectively, with $N_A+N_B=N$.
The atomic fractions will be denoted as $x$ for $B$ and $1-x$ for $A$, such that $N_B = N x$ and $N_A = N (1-x)$.
Now, if all sites are occupied independently by the $N$ atomic nuclei, they can be arranged in a total of
\begin{equation}
    W_\text{point} = \frac{N!}{N_A! N_B!}
\end{equation}
ways.
Points, however, cannot be occupied independently unless we have an ideal solution. 
The GQCA accounts for correlations between atoms by treating the lattice as an ensemble of $M = qN$ clusters of atoms and correcting for the overcounting of lattice points shared between clusters.
We will consider that each cluster is a geometric figure (a pair, a triangle, a cube, etc.) consisting of $n$ lattice sites. All clusters have the same geometry, differing only by their site occupations. A key factor is that the clusters overlap with each other. The factor $q$ is the  \emph{cluster coordination number}, or the number of clusters per lattice point. For instance, for a 2D square lattice, and considering the nearest-neighbor (NN) pair as cluster, $q=2$, and for an fcc lattice, still using the NN pair as cluster, $q=6$. Now, considering for a moment that all $M$ clusters are independent, the number of ways of arranging them throughout the lattice is
\begin{equation}
    W_\text{cluster} = \frac{M!}{\prod_j M_j!} \,,
    \label{eq:Wcluster-non-deg}
\end{equation}
in which $M_j$ is the number of clusters of type $j$. By cluster type we mean clusters with the same composition but different site occupations. For instance, if the cluster is a NN pair in a binary $A$--$B$ alloy, there will by four types of clusters: $AA$, $AB$, $BA$ and $BB$. In this way, we are considering that the pairs $AB$ and $BA$ are different (i.e., non-degenerate). If we consider that some pairs are energetically degenerate and symmetrically equivalent, we can in principle separate the pairs by classes and rewrite $W_{cluster}$ as 
\begin{equation}
    W_\text{cluster} = \frac{M!}{\prod_j M_j!} \prod_j (g_j)^{M_j} \,,
    \label{eq:Wcluster-deg}
\end{equation}
because, once a cluster is assigned to a class, there are $g_j$ ways to distribute the atomic species between its $n$ points, and there are $M_j$ clusters in class $j$. It should be noted that, in Eq.~\eqref{eq:Wcluster-non-deg}, the index $j$ runs through all types of (non-degenerate) clusters, while, in Eq.~\eqref{eq:Wcluster-deg}, the index $j$ runs through all classes of degenerate clusters. In the binary example used previously, using the NN pair as cluster, we would have three classes: $AA$, $AB$ and $BB$, with degeneracies $1$, $2$, and $1$, respectively. Since we are dealing with a binomial distribution, the degeneracies follow the relationship
\begin{equation}
    \sum_j g_j = 2^n \,.
\end{equation}
The $M$ clusters are not independent, though, because they overlap at lattice sites. Eq.~\eqref{eq:Wcluster-deg} must be duly corrected to take that into account. Because the clusters overlap, each lattice site is shared among $n q$ clusters, and, therefore, from Eq.~\eqref{eq:Wcluster-deg} we must discount the configurations of $n q-1$ points from $W_\text{cluster}$, leaving the configurations for only one point in the final expression. This means that the total number of configurations of a distribution of $M$ overlapping clusters in the lattice will be given by
\begin{equation}
    W = \frac{W_\text{cluster}}{W_\text{point}^{n q-1}} \,.
    \label{eq:W-GQCA-def}
\end{equation}
Other corrections, such as triplets, quadruplets, and other subclusters should also be included to take multi-site correlations into account. When corrections for all subclusters belonging to the maximal cluster are taken into account, we arrive at the Cluster Variation Method (CVM) entropy expression. The GQCA includes only point corrections. From Eq.~\eqref{eq:W-GQCA-def}, and using Eqs.~\eqref{eq:Wcluster-non-deg} and \eqref{eq:Wcluster-deg}, this leads to
\begin{equation}
    W = \frac{M!}{\prod_j M_j!} \prod_j (g_j)^{M_j}
        \left( \frac{N!}{N_A! N_B!}\right)^{1-n q}
\end{equation}
Introducing now the cluster class probabilities $p_j$, given by
\begin{equation}
    p_j = \frac{M_j}{M} \,,
\end{equation}
with
\begin{equation}
    \sum_j p_j = 1 \,,
\end{equation}
and using Stirling's approximation
\begin{equation}
    \ln u! = u \ln u - u + \mathcal{O}(\ln u)
\end{equation}
in Boltzmann equation for the entropy, we arrive at
\begin{align}
    S = & - k_B M \sum_j p_j \ln \frac{p_j}{g_j} \nonumber \\
       & - k_B N (1-n q) [x \ln x + (1-x) \ln (1-x)] \,,
    \label{eq:S-degen}
\end{align}
where $j$ runs through all cluster classes. 
Eq.~\eqref{eq:S-degen}, can be modified by introducing the random class cluster probabilities $p_j^0$, given by
\begin{equation}
    p_j^0 = g_j {1-x}^{n-n_j} {x}^{n_j} \,,
    \label{eq:p0j-def}
\end{equation}
in which $n_j$ is the number of $B$ nuclei in a cluster of class $j$, with $0 \le n_j \le n$. Eq.~\eqref{eq:p0j-def} gives the probability of randomly placing $n_j$ nuclei of type $B$ and $n-n_j$ of type $A$ in a cluster of class $j$ with degeneracy $g_j$. These are also called ``a priori'' cluster probabilities. One constraint comes from the balance on the total number of each species:
\begin{equation}
    \sum_j p_j n_j= n x
    \label{eq:Nk const}
\end{equation}
(with a similar equation for $A$), which simply states that the total number of nuclei of any type in the lattice must remain constant. Isolating $g_j$ from Eq.~\eqref{eq:p0j-def} and substituting it into Eq.~\eqref{eq:S-degen}, together with Eq.~\eqref{eq:Nk const}, leads, after simplification using $M = N q$, to
\begin{equation}
    S = -k_B M \sum_j p_j \ln \frac{p_j}{p_j^0} - k_B N [x \ln x + (1-x) \ln (1-x)]
    \label{eq:S-GQCA}
\end{equation}
Eq.~\eqref{eq:S-GQCA} is the usual form for the entropy in GQCA, being equivalent to Eq.~\eqref{eq:S_GQCA}. It can be readily seen that Eq.~\eqref{eq:W-GQCA} will result in the same expression for the entropy.

In general, because the clusters overlap throughout the lattice, each point takes part in $n q$ clusters (remember that $q$ is the cluster coordination number, i.e., the number of clusters per lattice point in the lattice, and $n$ is the number of points in a cluster). On the other hand, as stated in the main text, if a stacking of unit cells (that is, a supercell) is used as cluster, and the clusters overlap with each other at lattice sites, each point will be part of $n$ clusters. In this case, therefore, we have $nq=n$, or $q=1$, which means that $M = N$. For other kinds of cluster, in particular for smaller clusters, such as pairs or tetrahedra, $q$ will not be, in general, equal to unity, but, as always, it will come from the geometry of the lattice.



\section*{Author Contributions}

PNF conceived the work.
DS, LW, LTFE, and PNF developed the code.
LTFE revised the thermodynamic fundamentals and derivations.
DS and LW ran the EGQCA benchmark.
DS and PNF performed the DFT calculations. 
KB trained the EDDPs.
RM assisted with the ML models.
DS, LW, and PNF wrote the first draft.
SDC, EM, EK, CP, MA, CH participated in the discussions.
PNF supervised this project. 
All authors revised the manuscript.

\section*{Declaration of competing interest}
All authors declare no financial or non-financial competing interests.

\section*{Acknowledgments}

We thank Lara K\"uhl Teles and Marcelo Marques for insightful discussions. PNF acknowledges support from the Austrian Science Fund (FWF) under project DOI 10.55776/ESP8588124. DS, LW, and CH acknowledge support from the Austrian Science Fund (FWF) under project DOI 10.55776/PAT1157425. KB acknowledges support from the UK Engineering and Physical Sciences Research Council [EP/W524633/1].
The computational results have been achieved using local workstations at TU Graz, Austrian Scientific Computing (ASC) infrastructure, and LUMI. KB acknowledges the EuroHPC Joint Undertaking for awarding this project access to the EuroHPC supercomputer LUMI, hosted by CSC (Finland) and the LUMI consortium through a EuroHPC Regular Access call.

\section*{Data availability}
All data used to prepare this manuscript are freely available on Zenodo (\url{https://doi.org/10.5281/zenodo.24534160}).

\section*{Code availability}
The underlying code for this study is available on github (\url{https://github.com/egqca/egqca}) and as python package \texttt{egqca} (\url{https://pypi.org/project/egqca/}).

\section*{Declaration of generative AI and AI-assisted
technologies in the writing process}
In preparing this paper, the authors used large language models to help improve readability and language. After using these tools, the authors reviewed and edited the content as needed and take full responsibility for the content of the published article.


\bibliographystyle{apsrev4-2}
\bibliography{refs}

\clearpage
\newpage
\begin{figure*}
\includegraphics[page = 1, width=\linewidth]{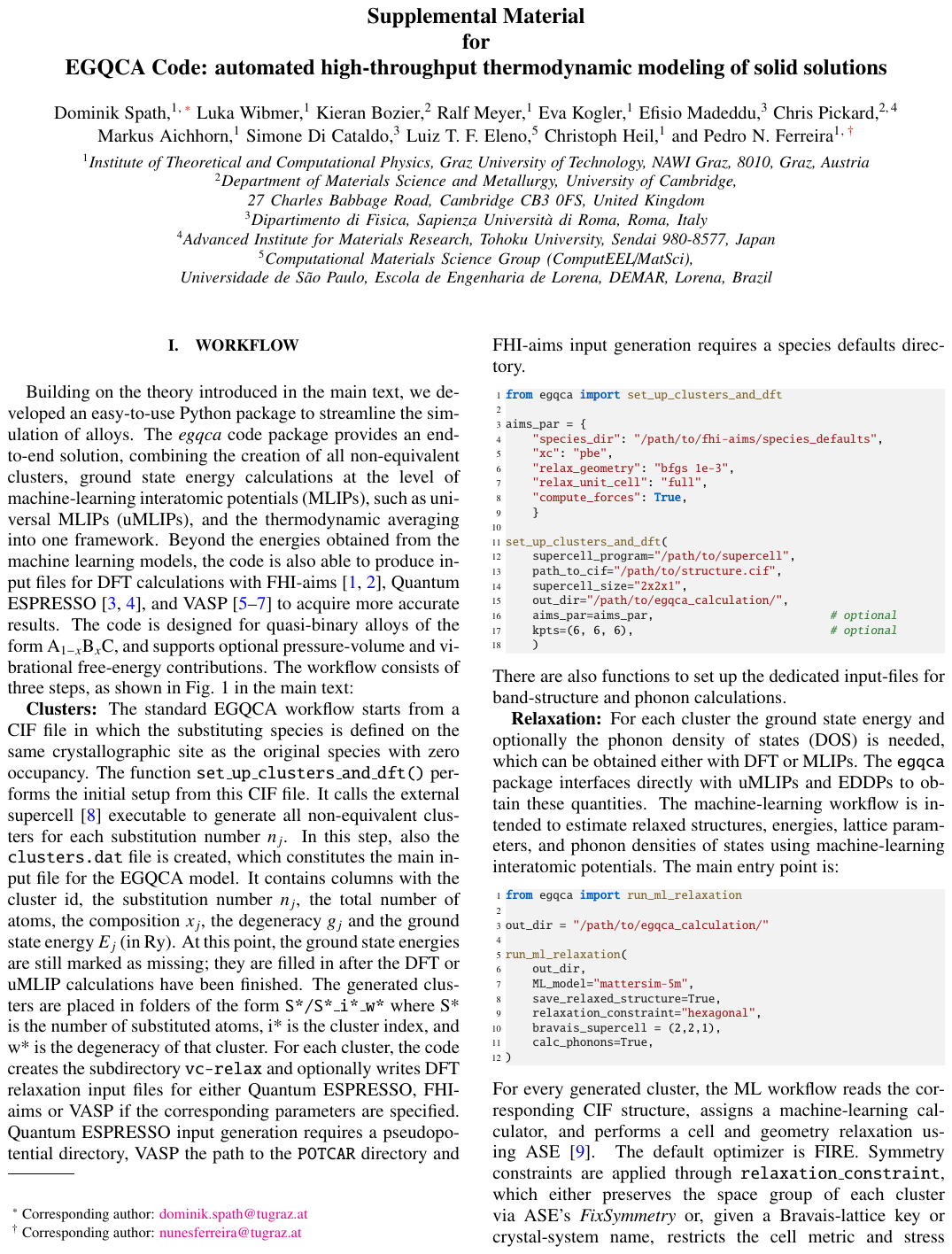}
\end{figure*}
\begin{figure*}
\includegraphics[page = 2, width=\linewidth]{supplemental.pdf}
\end{figure*}
\begin{figure*}
\includegraphics[page = 3, width=\linewidth]{supplemental.pdf}
\end{figure*}
\begin{figure*}
\includegraphics[page = 4, width=\linewidth]{supplemental.pdf}
\end{figure*}
\begin{figure*}
\includegraphics[page = 5, width=\linewidth]{supplemental.pdf}
\end{figure*}

\end{document}